\documentclass[namedreferences]{kluwer}

\usepackage{graphics,psfig,rotate}

\begin{document}
\begin{article}
\begin{opening}

\title{Scaling Laws in Self-Gravitating Disks}

\author{Daniel \surname{Huber} \& Daniel \surname{Pfenniger}}

\institute{Geneva Observatory, Ch. des Maillettes 51, CH-1290 Sauverny,
           Switzerland}

\date{Received / Accepted}

\begin{abstract}

The interstellar medium (ISM) reveals strongly inhomogeneous structures
at every scale.  These structures do not seem completely random since
they obey certain power laws. Larson's law (\citeyear{Larson81}) $\sigma
\propto R^{\delta}$ and the plausible assumption of virial equilibrium
justify to consider fractals as a possible description of the ISM.  In
the following we investigate how self-gravitation, differential
rotation and dissipation affect the matter distribution in galaxies.
To this end we have performed 3D-simulations 
for self-gravitating local boxes embedded in a larger
disk, extending the 2D-method of Toomre \& Kalnajs 
(\citeyear{Toomre91}) and Wisdom \& Tremaine
(\citeyear{Wisdom88}). Our simulations lead to 
the conclusion that 
gravitation, shearing and dissipation can be dominantly responsible
for maintaining an inhomogeneous and eventually a fractal distribution
of the matter. 

\end{abstract}

\keywords{Methods: numerical - Galaxies: ISM - ISM: structure}

\end{opening}

\section{Introduction}

The lumpy distribution of matter resulting in Toomre \& Kalnajs'
(1991, hereafter TK) local shearing-sheet experiments, reminds us of
the ubiquitous inhomogeneous state of the ISM as well as the flocculent
structures of many spirals.  Therefore we decided to take up their
2D-model and to extend it on 3 dimensions 
order to investigate more precisely the clumpy structure of galactic
matter. It is important to include the third dimension in the model,
because as soon as clumping develops dynamical coupling with vertical
motion must be strong, contrary to the weak coupling existing in a smoothly
distributed thin disk. Our model considers scales of the order of
${\cal O}(1 \rm{kpc})$. Thus we are able to
investigate the transition 
regime between the molecular cloud scales and the galactic disk scale.
Furthermore we can investigate whether gravitation and dissipation can maintain
the power laws observed in molecular clouds on scales at which
shearing is dominant. 

This paper is divided up in two main parts. The first part presents the 
model (Sects. 2.1-2.4) and the second part shows the results (Sect. 3.1)
and their analysis, i.e. the determination of the fractal dimension 
(Sect. 3.2) and the verification of Larson's law (Sect. 3.3).

\section{Model}
\subsection{Principle}

In local models only a small part (e.g., everything inside a box with a
given size) of the system is simulated and more distant regions are
represented by replicas of the local box.  In such a model 
the orbital motion of the particles is
determined by Hill's approximation of
Newton's equations of motion (Hill 1878)
\begin{equation}
\label{eq1}
\begin{array}{ccccccc}
\ddot{x}&-&2\Omega_0\dot{y}&=&4\Omega_0 A_0 x& + &F_x\\
\ddot{y}&+&2\Omega_0\dot{x}&=& & &F_y\\
\ddot{z}& &                &=&-\nu^2 z&+&F_z
\end{array}
\end{equation}
where $A_0=-\frac{1}{2} r_0 \left(\frac{d\Omega}{dr}\right)_{r_0}$ is
the Oort constant of differential rotation and $\nu$ is the vertical
epicycle frequency. $F_x, F_y$ and $F_z$ are
local forces due to the self-gravitating particles. 

Because of the shearing flow 
the relative positions of the rectangular
boxes (local box and replicas) change with time, so that a initially
periodic arrangement of the boxes relative to a fixed Cartesian
coordinate system can not be maintained.  As a consequence the forces
of the self-gravitating particles must be determined by direct
summation, requiring ${\cal O}(N^2)$ operations for $N$ particles.

To increase the performance in our models we calculate the forces with
the FFT-convolution method, requiring 
${\cal O}(N_c{\rm log} N_c)$ operations, where $N_c$  is
the number of cells, which should have the same order of magnitude as
the number of particles. This requires a system spatially
periodic at each time.
Therefore we use a pair of time-dependent affine coordinate systems, whose
affinity angle change with time and are determined by the shear flow (see
Fig. 1). The angle
difference between the two grids is constant and
given by ${\rm atan}(L_y/L_x)$.  The reason to evaluate twice the
forces in two different affine grids is for avoiding force
discontinuities in time when the affine angle would have reached a
maximum value beyond which a smaller angle would exist. After the
calculation of the forces for the two affined grids, they are
transformed into Cartesian coordinates, where they are weighted and
added. The weighting factors are proportional to the grid inclination 
and normalized.

\begin{figure}[htb]
\centerline{
\psfig{file=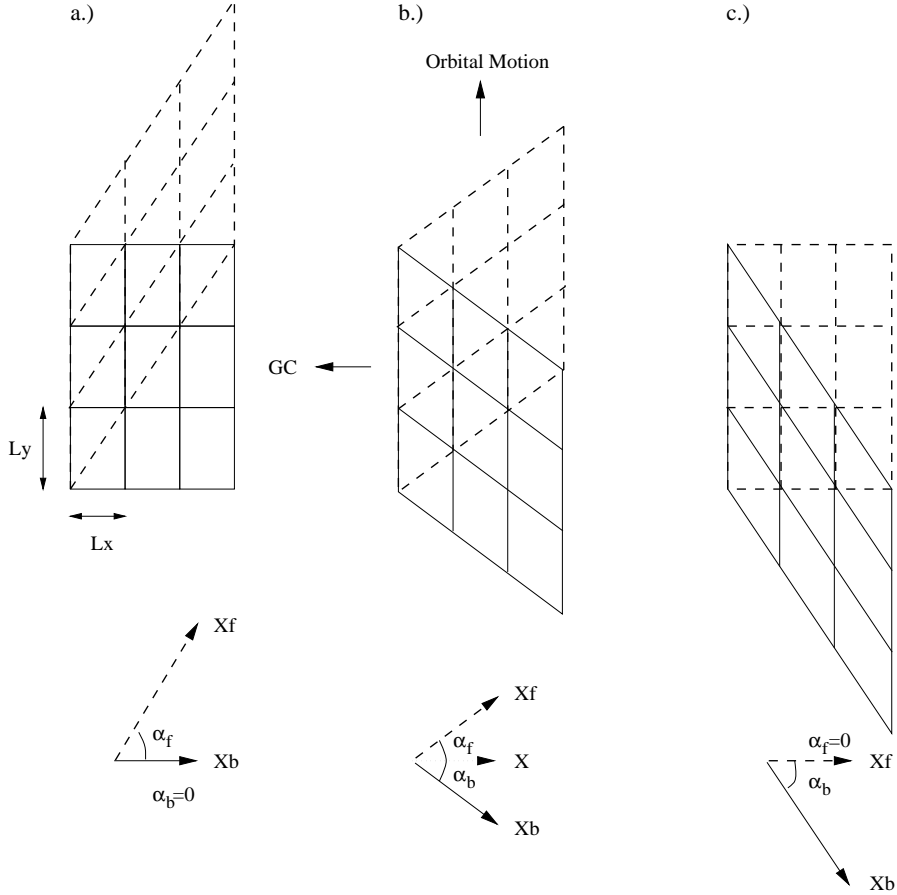,width=\hsize}}
  \caption{Schematic diagram of the inclined grids seen 
  from above. {\it a.)}  The initial state of the two grids $(t=0)$. The
  dashed grid is the forward grid and the solid one is the backward
  grid. Below them, the affinity angles of the grids are indicated, 
  $\alpha_f$ resp. $\alpha_b$.
  The angle between the two grids remains the
  same for all times. {\it b.)} The grids at $t=L_y/(2
  L_x \Omega_0)$, where $\alpha_f=-\alpha_b$. 
  {\it c.)} When the grids attain these
  inclinations, they jump back to the positions shown in {\it (a)} and
  the process starts again without discontinuity in the dynamics.}
\end{figure} 


\subsection{Cooling}
\label{cooling}
To avoid an increase of the random epicyclic motion due to
gravitational heating, TK proposed to add artificial cooling
forces. Following them we include the damping terms $-C_x \dot{x}$ and
$-C_z \dot{z}$  in the radial resp. vertical forces $(F_x, F_z)$
controlling the particle motions via Eq. (\ref{eq1}). Two different
damping terms are necessary to reflect the different directional collision
rates induced by the anisotropic velocity ellipsoid. The cooling
coefficients are chosen in such a way that the velocity dispersions of the
random motions reach a stable level and don't differ much from their 
initial values. The same stability conditions holds for the disk
scale height $z_0$.

\section{Results}
\subsection{Shearing boxes}

\begin{figure*}
\psfig{file=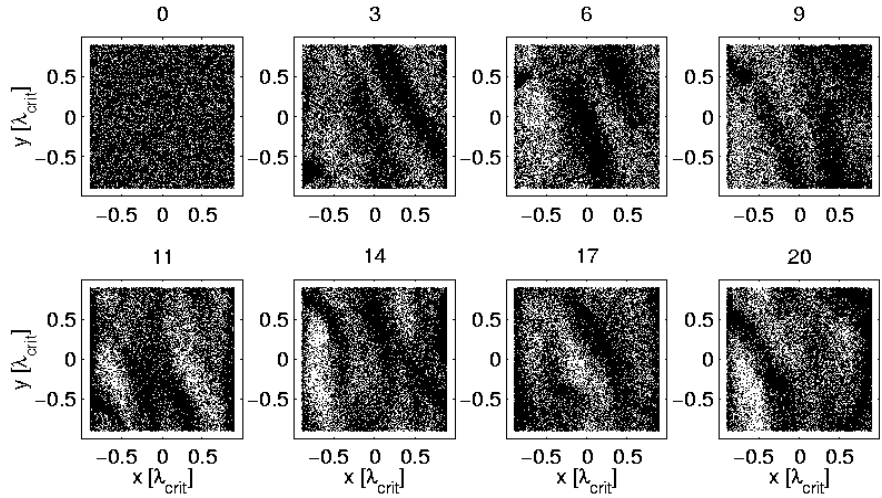,angle=90,width=\hsize}
\caption{ The evolution of the particle positions seen
    from above. The number of rotations of the shearing box around
    the galactic center are indicated above each panel. The number
    density at the beginning of the simulation is $n=10100
    \lambda_{\rm crit}^{-2}$, corresponding to $32724$ particles.}
\label{xy}
\end{figure*}

\begin{figure*}
\psfig{file=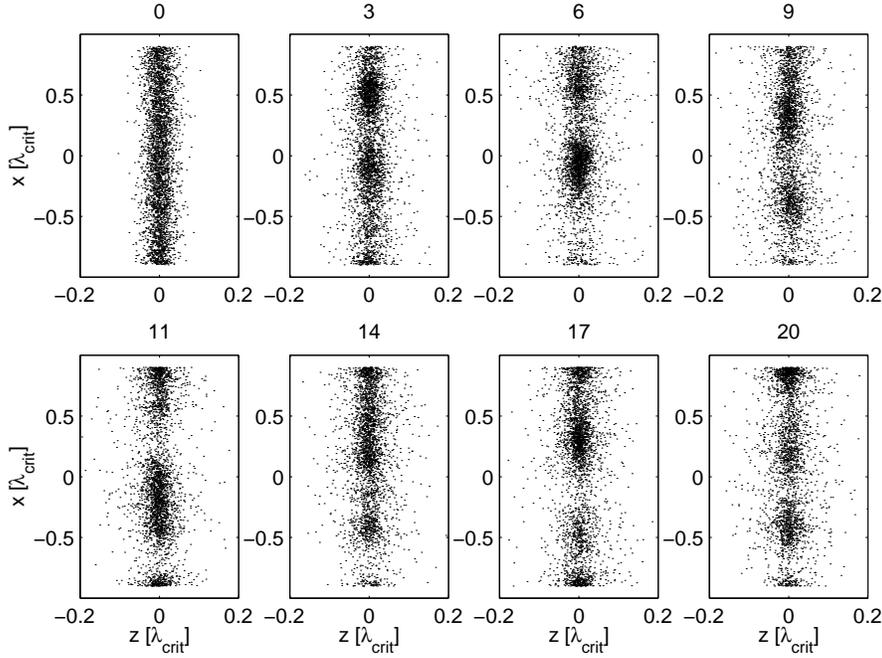,angle=90,width=\hsize}
\caption{The positions of the particles inside the slice,
    $-0.1\; \lambda_{\rm crit}< y < 0.1\; \lambda_{\rm crit}$, seen in the
    direction of orbital motion.}
\label{xz} 
\end{figure*}

Fig. \ref{xy}\footnote{Full resolution paper available at
      http://obswww.unige.ch/Preprints/cgi-bin/Preprintshtml.cgi?\#DYNAMIC.}
and \ref{xz} reveals the spatial distribution of matter in a
box, representing a section of a galactic disk. The size of the box
sides $(L_x\times L_y\times L_z)$ is given by the critical wavelength 
$\lambda_{\rm crit}$, defining the scale for which the theory of swing 
amplification predicts the strongest response
(Toomre \citeyear{Toomre81}).

Because we are interested in long time behavior, the simulations were
performed for $t=20$ galactic rotations. In the initial state $t=0$
the particles are distributed uniformly in the x-y-plane. The particle
distribution in $z$-direction obeys $\rho\propto{\rm sech}^2(z/z_0)$, 
where $\rho$ is the density and $z_0$ is the disk scale height.
The velocities at $t=0$ are determined by the shear-flow
\begin{equation}
\begin{array}{ccccccccccc}
\dot{x}&=&0,& &\dot{y}&=&-2A_0 x,& &\dot{z}&=&0
\end{array}
\end{equation}
and the Schwarzschild velocity ellipsoid.

Like the 2D-model of TK, our 3D-model reveals a fast ``structure
formation''. After the first rotations the striations are already developed
and are more or less maintained during the rest of the simulation.

In our model 
$t_{\rm osc} < t_{{\rm cool},x} < t_{{\rm cool},z}$ is valid, where
$t_{\rm osc}$ is the period of the unforced epicyclic motion. 
$t_{{\rm cool},x}$ and $t_{{\rm cool},z}$ 
are the cooling times for the radial resp. vertical damping in fact
$t_{cool,x} \propto C_x^{-1}$ and $t_{cool,z} \propto C_y^{-1}$. 
Furthermore, the 
cooling times depend
on the particle number density $n$. For the simulation in Fig. \ref{xy}
the following relation is valid: 
$t_{\rm osc} : t_{{\rm cool},x} : t_{{\rm cool},z} \approx 1 : 40 : 300$. 

\begin{figure}[htb]
\centerline{
\psfig{file=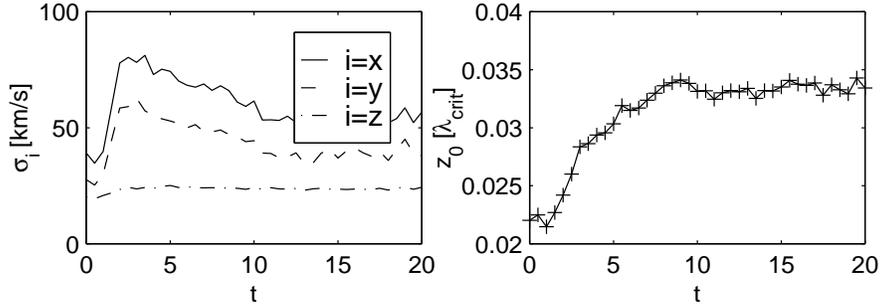,angle=90,width=\hsize}}
\caption{The left panel shows the velocity 
  dispersions $\sigma_x$, $\sigma_y$
  and $\sigma_z$ as a function of time $t$, indicated in galactic
  rotation unit. The right panel reveals the evolution of the disk 
  scale height $z_0$.}
\label{dispz0}
\end{figure}

The random velocity dispersions and the disk scale height were 
calculated after every half rotation and are plotted in
Fig. \ref{dispz0}. Gravitational 
instability and shearing, which are responsible for an important part 
of the striations, have an effect particularly in the x-y-plane and leads to 
a fast increase of $\sigma_x$ and $\sigma_y$. 

\subsection{Fractal dimension}
During this studies our interest was particularly focused on the
following question: Can self-gravitation, shearing and dissipation
be dominantly responsible for maintaining a fractal distribution of the
matter in a disk and can they account for Larson's law (Larson 1981)?
To answer this question we calculated the appropriate scaling laws and 
determined the
fractal dimension for the structures shown in
Fig. \ref{xy} and \ref{xz}. 

\begin{figure}
\centerline{
\psfig{file=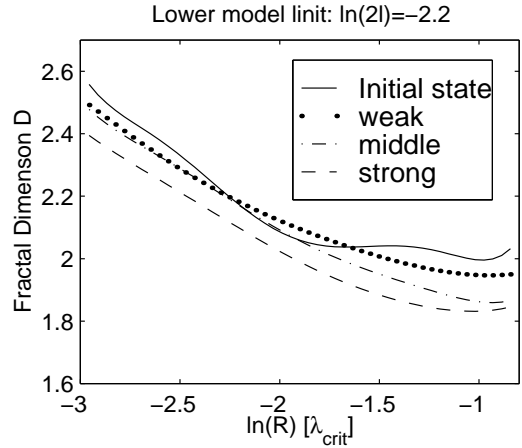,width=7cm}}
\caption{ The structure dimension $D$ as a function
    of the scale $R$. The solid line corresponds to the initial
    state. On large scales this state represents a 2D matter
    distribution, whereas on small scales $R<z_0$ it tends to $D>2.6$.
    The other curves represent the simulation terminal phase for
    different cooling forces (weak, middle, strong). The
    relative magnitudes of the cooling coefficients are: 
    $C_{x,{\rm weak}} : C_{x,{\rm middle}} : C_{x,{\rm strong}}\approx
    1:1.5:2$. Above the figure the natural logarithm of the lower model 
    limit is indicated.}
\label{fractal3}
\end{figure}

We determined the fractal dimension, quantifying how much space our
system fills, as follows: We choose a representative set of particles
and count for each particle the number of neighboring particles
$N(R)$ inside a certain radius $R$. If we repeat this for other values
of $R$ we can find the fractal dimension $D$ via
\begin{equation}
N(R)\propto R^D.
\end{equation}
Because the structures we examine are not an idealized mathematical
set, but model a physical system, we have to take into account 
upper and lower cutoffs in the 
analysis.  An upper limit due to the
numerical model is given by the size of the simulation box in the
x-y-plane. On this scale the system becomes periodic, meaning that it
can not be fractal on scales ${\cal O}(L_x)$.
 A lower limit is due to the finite resolution of
the box grid. If the grid cells have the size $l_x\times l_y\times l_z$ 
and $l=l_x=l_y>l_z$ then we can't expect fractal structures below $2 l$.
Fig. \ref{fractal3} reveals the fractal dimension 
$D=d {\rm ln}(N)/d{\rm ln}(R)$
as a function of the radius $R$, i.e. of the scale. The solid line
corresponds to the initial distribution of matter: 
$\rho(x,y)={\rm const.},\; \rho(z)\propto{\rm sech}^2(z/z_0)$. The other
lines give the mean dimensions of the structures for the simulation
terminal phase. The intensity of the structures and thus the
structure dimension $D$ depend on the cooling. To show this clearly we 
performed simulations with varying cooling forces. The figure reveals
clearly that the stronger the cooling, the more intensive the
structures and the lower the structure dimension $D$ are. The
structures in Fig. \ref{xy} and \ref{xz} were performed with a ``strong'' 
cooling. For this structures $D$ runs in a thin band between 
$0.2\lambda_{\rm crit}$ and the upper model limit with a minimum of  
$D\approx 1.83$ at $R=0.35 \lambda_{\rm crit}$. Pfenniger 
(\citeyear{Pfenniger96}) showed, that a fractal-like,
hierarchical gravitating system in statistical equilibrium can only
exist with $D<2$. In the range, in which $D$ runs in a thin band this
condition is fulfilled, but the dynamical range is too small to call
the structures fractals.

\subsection{Larson's law}
If the fractal organization of the matter extends to scales at which
the shear flow becomes important and the system is still virialized,
then the validity range of Larson's law
\begin{equation}
\sigma \propto R^{\delta} \quad \hbox{\rm or} \quad
\delta=\frac{d\ln \sigma}{d\ln R}
\end{equation}
should exceed the size of
Giant Molecular Clouds. 
We checked Larson's law for scales $>100$ pc
by analyzing the phase-space of the particles in Fig. \ref{xy}.  

\begin{figure}
\centerline{
\psfig{file=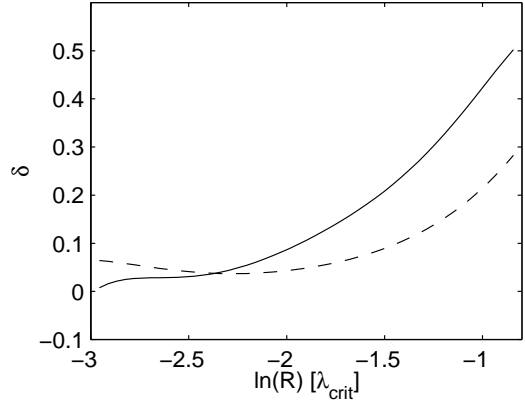,width=7cm}}
\caption{$\delta$ of Larson's law as a function
    of the scale $R$. To calculate $\delta$ we used the simulation
    with the particle density $n=40450$ and the resolution 
    $l=0.03 \lambda_{\rm crit}$ The solid line corresponds to the initial
    state, whereas the dashed line indicate $\delta$
    for the simulation terminal phase.}
\label{larson}
\end{figure}

Fig. \ref{larson} shows $\delta$ as a
function of the scale $R$.
There is no plateau between $0.3<\delta<0.5$, but $\delta$ reaches this
range for the bigger scales, where also $2\delta+1= D$ is roughly
fulfilled. 

\section{Conclusion} 

Our local 3D-simulations show, that self-gravitation and dissipation 
ensure a statistical equilibrium at scales at which the shear flow is 
important. Moreover they reveal a fractal-like distribution of matter 
on kpc-scales. The fractal dimension found in the simulations can be 
lower than 2, as observed in molecular clouds. The specific value
found depends on the relative strengths of the competing gravitational 
and dissipation processes. Because the dynamical range of our
simulations is still small, it would be premature to call the found
structures strict fractals. The anisotropy of the velocity-dispersion
ellipsoid, resulting from our simulations, has systematically the same 
ordering $(\sigma_R>\sigma_\Phi>\sigma_z)$ as observed in the Galaxy
and in N-body simulations of spirals.

\begin{acknowledgements}

This work has been supported by the Swiss Science Foundation.

\end{acknowledgements}

\end{article}
\end{document}